\documentclass{article}

\usepackage{graphics,amsfonts}

\begin{document}
\begin{titlepage}
\hskip 12cm
\vbox{\hbox{hep-th/9708164}\hbox{CERN-TH/97-223}\hbox{September, 1997}}
\vfill
\begin{center}
{\LARGE {Duality, Strings and Supergravity: a Status Report }}\\
\vskip 1.5cm
{ 
Sergio Ferrara } \\
\vskip 0.5cm
{\small
 CERN Theoretical Division, CH 1211 Geneva 23, Switzerland
}
\end{center}
\vfill
\vskip 3cm
\begin{center}
Plenary Talk given at the\\
 XII International Congress of Mathematical Physics\\
13-19 July 1997,  Brisbane, Australia
\end{center}
\vskip 3cm
\begin{center} {\bf Abstract}
\end{center}
{
\small
We will report on recent advances in the understanding of non-perturbative
interconnections between different string dualities.
Weak-strong coupling duality (S-duality) and T-duality (symmetry under
 compactification on dual tori)
allows one to compare  and explore the strong coupling regime of seemingly
 unrelated theories.
These theories naturally merge in a quantum version of supergravity called
 M-theory.
The dynamical role of `branes' of different nature and the new dynamical tool
 of 
(M)atrix formulation of M-theory will be briefly mentioned.
}

\vspace{2mm} \vfill \hrule width 3.cm

\end{titlepage}

\title{Duality, Strings and Supergravity: a Status Report}

\author{ Sergio FERRARA %
  \footnote{Theoretical Physics Division, CERN, 1211 Geneva 23, Switzerland,
            email: \texttt{ferraras@vxcern.cern.ch}.}
}

\maketitle

\font\cmss=cmss10 \font\cmsss=cmss10 at 7pt
\def\IR{\relax{\rm I\kern-.18em R}}
\def\ZZ{\relax\ifmmode\mathchoice
{\hbox{\cmss Z\kern-.4em Z}}{\hbox{\cmss Z\kern-.4em Z}}
{\lower.9pt\hbox{\cmsss Z\kern-.4em Z}}
{\lower1.2pt\hbox{\cmsss Z\kern-.4em Z}}\else{\cmss Z\kern-.4em
Z}\fi}
\def\bfone{\relax{\rm 1\kern-.35em 1}}

\abstract{We will report on recent advances in the understanding of
 non-perturbative
interconnections between different string dualities.
Weak-strong coupling duality(S-duality)  and T-duality (symmetry under
 compactification on dual tori)
allows one to compare  and explore the strong coupling regime of seemingly
 unrelated theories.
These theories naturally merge in a quantum version of supergravity called
 M-theory.
The dynamical role of `branes' of different nature and the new dynamical tool
 of 
(M)atrix formulation of M-theory will be briefly mentioned. }


\section{Introduction}
In the past three years, since the last ICMP held in Paris in 1994,
stunning developments in non-perturbative quantum theories of basic forces
 have been
taking place.
These dramatic results occurred in two different but related types of
quantum theories:
supersymmetric Yang--Mills theories in four dimensions, \cite{aaa} and
superstrings \cite{bb}.
These theories have in common two basic symmetries:
1) space-time supergravity, which controls both perturbative and
non-perturbative effects,
2) ``duality'', which allows  relating different (or the same)
theories in different coupling constant regimes and fundamental states
to solitonic excitations.

  These theories also share the property that a
formal perturbative expansion can be defined for them, and
``renormalization" can be used in order to extract finite answers from any
perturbative calculation.

They enjoy non-renormalization theorems~\cite{cc} depending on the degree
of supersymmetry of the vacuum around which we define the quantum perturbative
series.

Let us recall the spin content of the light states:
the gauge theory  includes massless quanta with spin 0, 1/2, 1;
superstrings include massless quanta with spin 0, 1/2, 1, 3/2 and 2; in
a suitable limit ($\alpha' \rightarrow 0$, fixed gauge-coupling)
superstrings reproduce ordinary gauge theories.

Exact non-perturbative results of SQCD have been obtained.  Seiberg and
Witten~\cite{dd} suggested an exact expression for the low-energy effective
 action
for the Coulomb phase of an $N = 2$ SU(2) SYM theory, which may be regarded as
 an
extension of the Georgi--Glashow SU(2) gauge theory.

Electric--magnetic duality plays a crucial role to solve the theory, to compute
strong coupling phases of this theory (where massless monopoles and dyons
appear), and to prove colour confinement through a magnetic Higgs mechanism
 with
a monopole condensation (analogous to Meissner effects in superconductors).

The solution of the problem is possible thanks to non-renormalization theorems,
making the complete perturbative computation affordable.

The non-perturbative part, conjecturally due to processes from multi-instanton
transitions, is obtained from a mathematical hypothesis that identifies
 electric
and magnetic massive states with windings $(n,m)$ of a genus 1 (torus)
elliptic Riemann surface (genus $n$ for an SU($n$) gauge-theory as shown by
 Klemm,
Lerche, Theisen and Yankielowicz and by Argyres and Faraggi~\cite{ee}).
Two topologically different cycles correspond to electric and magnetic charges.
Quantum massive BPS states $\psi(n,m)$ correspond to distinct topological
configurations $(n,m)$ of the elliptic surface, in particular
$\psi(1,0)=W$-boson, $\psi(0,1) =$ monopole, $\psi(-1,+1)$ = dyon.


\section{Electric--Magnetic Duality and Supersymmetry}
Let us recall the duality of Maxwell equations
\begin{eqnarray}
&\nabla \cdot (E + iB) = \rho_e + i\rho_m = \rho \nonumber\\
&\nabla \wedge (E + iB) - i \frac{\partial}{\partial t} (E + iB) = J_e + iJ_m =
J\nonumber\\
& L = {\rm Re}(E + iB) \cdot (E + iB) = E^2 - B^2~.\nonumber
\end{eqnarray}
Here $E, B$ are the electric and magnetic fields;  $\rho_e \, (\rho_m), J_e \,
 (J_m)$
denote electric (magnetic) charge density and current, respectively.
 A topological
term $E \cdot B$ may eventually be added to $L$.  The physical observables
 such as
the energy density $(E + iB) \cdot (E - iB) = E^2 + B^2,$ and the momentum
 density
$(E + iB) \wedge (E - iB)$ are invariant under (continuous) U(1) duality
 rotations:
\[
(E + iB) \rightarrow e^{i\varphi}(E + iB)~, ~~ \rho \rightarrow
e^{i\varphi}\rho~,~~J \rightarrow e^{i\varphi}J~.
\]
In particular the $Z_2$ symmetry, which is the remnant of U(1), acting on
 discrete
charged states, exchanges electric with magnetic fields $E \rightarrow B~{\rm
and}~B \rightarrow -E,$ and electric and magnetic charges $q
\rightarrow g, g \rightarrow -q$, accordingly.

The simultaneous occurrence of electric and magnetic sources implies a charge
quantization, which reads:
\begin{eqnarray}
&qg = 2\pi k\hfill \nonumber \\
&{\rm\hbox{ (Dirac~1931)~\cite{ff}}}\, {\rm (for~monopoles)}\nonumber
\end{eqnarray}
and
\begin{eqnarray}
&q_1g_2 - q_2g_1 = 2\pi k \nonumber \\
&{\rm
\hbox{(Schwinger,~Zwanziger~1968)~\cite{ggg}}}\,
 {\rm (for~dyons).}\nonumber
\end{eqnarray}
In the Coulomb phase the Georgi--Glashow SU(2) gauge theory has a monopole with
mass ('t~Hooft, Polyakov 1974)~\cite{hh}:
\begin{eqnarray}
&M_{{\rm monopole}} \geq 1/\lambda \langle \phi \rangle , \nonumber \\
&{\rm\hbox{
(Bogomolny~bound, ~1975)~\cite{jj}}}\nonumber
\end{eqnarray}
while the classical vector-boson mass is $M_W = \lambda \langle \phi \rangle$.
In the Prasad--Sommerfeld (1976) limit~\cite{kk} (supersymmetry),
$M_{\rm monopole} = 1/\lambda \langle \phi \rangle$
satisfies the duality conjecture (Montonen, Olive 1977)~\cite{lll}:
\[
M^2(q,g) = M^2(q^2 + g^2) = \langle \phi \rangle^2 (q^2 + g^2)~.
\]
This generalizes when a topological term $\theta E\cdot B$ is included by
 defining
a complex parameter $\tau = \theta + i/\lambda^2$ and then writing:
\[
M^2(\phi,\tau,n,m) = \frac{|\phi^2|}{lm\tau} |n + \tau m|^2
\]
invariant under $SL(2,Z)$:
\begin{eqnarray}
Z_2 &:& \tau \rightarrow \frac{-1}{\tau}~,~~n \rightarrow m~, ~~m \rightarrow
 -n
\nonumber\\
\theta-{\rm shift}&:& \tau \rightarrow \tau +1~,~~n\rightarrow n-m~,~~ m
\rightarrow m~.\nonumber
\end{eqnarray}
This means that the dual theory obtained by a $Z_2$ symmetry $E \rightarrow B,
 B
\rightarrow -E$ has $\tau_d = -1/\tau, n_D = m, m_D = -n$.
The $N=4$ supersymmetric Yang--Mills theory realizes the Montonen--Olive 
duality
conjecture~\cite{lll}.  The theory has an exact
SL(2,$Z$) symmetry~\cite{dda}, \cite{dd}, which is possible in virtue of a
vanishing
$\beta$ function, in the full quantum theory.
Electric states are fundamental, while magnetic states are solitons in the
 theory
$T$, but their role is reversed in the dual theory $T_D$ \cite{olive}.

Seiberg and Witten~\cite{dd} extended the duality to $N=2$, SYM quantum field
theories undergoing renormalization $(\beta \not= 0)$, which gives corrections
 to
a `holomorphic prepotential', $F(\phi)$;  this is the appropriate tool to build
up
$N=2$ effective actions.  The BPS states (which lie in hypermultiplets) have
 mass
$M(\phi,n,m,\lambda) \propto |\phi n +F_\phi m|$, where~\cite{mm}
$F(\phi) = (i/2\pi)\phi^2 \ln(\phi^2/\lambda^2) + \dots$ (the dots denote the
non-perturbative contributions).

They also extended the duality conjecture.  This came by identifying the pair
($\phi, F_\phi)$ with the periods of an hyper-elliptic surface, which allows us
to give a closed expansion for $F(\phi)$.  As a result, at strong coupling
$\phi^2/\lambda^2 =
\pm 1$, one gets a massless monopole (0,1) and a dyon ($-1,+1$).

The dual (U(1) magnetic) theory is weakly coupled in the strong coupling of the
electric theory and describes a magneto-dynamic of a charged monopole.  In the
weakly coupled magnetic Higgs phase, monopole condensation describes
 confinement of
the original (strongly coupled) electric theory.  It is worth mentioning that,
 for
BPS states,  their mass appears in the central extension of the supersymmetry
algebra (Haag--Lopuszanski--Sohnius)~\cite{nn} and this allows one, using
supersymmetry~\cite{oo}, to compute their mass in terms of the low-energy data.
The duality has been further extended to $N=1$ super-Yang--Mills
theories~\cite{pp}, in particular to SQCD with colour group SU$(N_c)$ and $N_f$
flavours.  This theory has an anomaly-free global symmetry:
\[
SU_L(N_f) \times SU_R(N_f) \times U(1)_B \times U(1)_R~.
\]
Seiberg suggested that there is a non-Abelian Coulomb phase for 
$3/2N_c < N_f <
3N_c$.  At the non-trivial infra-red fixed point, the theory of quarks and
 gluons
has a dual description in terms of an interacting conformal invariant theory 
with
magnetic gauge group $SU(N_f-N_c)$ and $N_f$ flavours.
Quarks and gluons are solitons in the dual picture.


\section{Supergravity, Strings and M-Theory}
Duality symmetries in the context of supergravity theories~\cite{qq}, further
extended to superstrings~\cite{rr}, allow us to prove exact equivalences of 
different
string theories~\cite{ss}--\cite{uu}, to obtain a dynamical
understanding of the Seiberg--Witten conjecture in the point-particles
limit~\cite{vv},\cite{ww} and finally to possibly merge these
theories in the context of M-theory, a supposedly existing quantum theory of
membranes and five-branes~\cite{yy}, \cite{zz}, whose low-energy effective
 action is
11-$D$ supergravity~\cite{aai}.

There are five known types of superstring theories in 10 dimensions~\cite{bb}:

\begin{tabular}{l | l }
Type & Gauge group\\ \hline
Type I & SO(32)\\
Heterotic & SO(32), $E_8 \times E_8$\\
Type IIA & U(1)\\
Type IIB & None
\end{tabular}

The first three have $N = 1$ supersymmetry, while the last two have $N = 2$,
non-chiral type IIA and chiral IIB. There is also a conjectured
M-theory in eleven dimensions~\cite{uu} (no gauge group).  Upon reduction on a
 circle,
this is equivalent to type IIA, at the non-perturbative level.  A further
speculative theory may exist in twelve dimensions, which gives, upon reduction
 on a
two-torus, the type IIB theory~\cite{bbi}.

The previous theories, and their compactification to lower dimensions, reduce
 at
low energies to supergravity theories in diverse dimensions~\cite{cci} with
underlying supersymmetry algebras as classified by Nahm~\cite{ddi}.  In the
highest and lowest dimensions of interest, we have for instance:
\begin{eqnarray}
D =   11, N &= &1, 128_{\rm boson} + 128_{\rm fermions}\nonumber\\
&&(b = 44 + 84, f =
128)\nonumber \\
D = 10, N & = &1~({\rm chiral})\nonumber \\
N &=& 1~({\rm matter}) (G = E_8 \times E_8, {\rm SO(32))}\nonumber \\
N &= & 2: {\rm\hbox{ type~A~(non-chiral)}},\nonumber \\
&& {\rm\hbox{type~B~(chiral)}}\nonumber\\
 D=4,  N&=&1~{\rm (chiral):~obtained~as}~M{\rm\hbox{-theory}}\nonumber\\
&&{\rm on}~M_7 = CY_3
\times S_1/Z_2 \nonumber\\
N&=&8~{\rm\hbox{(non-chiral)}}~(b = 56 + 70 + 2,\nonumber\\
&&f = 112 + 16):~{\rm obtained~as}~\nonumber \\
&& M{\rm\hbox{-theory on}}~T_7({\rm U}(1)^{28}~{\rm gauge}\nonumber\\
&& {\rm group)}\nonumber \\
&&{\rm or~on}~S_7({\rm SO}(8)~{\rm gauge~group)}~.\nonumber
\end{eqnarray}

Let us summarize some of the main
basic results of the years 94--97, in the context of string theory and
its non-perturbative regime.

\begin{itemize}
\item[1)] The Seiberg--Witten solution of rigid $N=2$ theory generalizes
to heterotic-type II duality relating $K_3 \times T_2$ vacua of heterotic
to Calabi--Yau vacua of type II strings~\cite{eei}, \cite{ffi}, \cite{kklmv},
 \cite{lerche}.

The second quantized mirror symmetry~\cite{eei} gives exact
non-perturbative results in
$N=2$ superstrings, $D = 4$.  In particular, duality relates world-sheet
instanton effects on the type II side to space-time instantons on the
heterotic side~\cite{vv}, \cite{ggi}.
Dual pair heterotic-type II theory constructions were proposed \cite{vafa}.

\item[2)] The implication of string--string duality in six dimensions for S--T
duality at $D = 4$ was first shown by Duff~\cite{ss}, and
$U$-duality as a non-perturbative symmetry of different string theories
was formulated by Hull and Townsend~\cite{tt}.

\item[3)] Witten~\cite{uu} proved the equivalence of different string
theories in higher dimension and the duality of type IIA at strong
coupling with 11-$D$ supergravity at large radius (M-theory on
$M_{10}\times S_1$).  Type IIB is self-dual at $D = 10$ (SL$(2,Z)$
duality)~\cite{hhia}.

\item[4)] The $E_8 \times E_8$ heterotic string at strong coupling is dual
to the M-theory on $M_{10} \times S_1/Z_2$ (Horava--Witten)~\cite{hhi}.

\item[5)] The SO(32) Type I and SO(32) heterotic at $D =
10$ are interchanged by  weak--strong coupling duality
(Polchinski--Witten)~\cite{jji}.

\item[6)] Open strings naturally arise, by the mechanism of tadpole
cancellations, as sectors of type IIB closed strings on
orientifolds~\cite{kki}, \cite{lli}.  Their end-points end on
$D$-branes~\cite{mmi}, carrying R--R charges.  Phase transitions in six
dimensions are possible~\cite{nni}, and evidence for a non-perturbative
origin of gauge symmetries~\cite{ooi} was substantiated~\cite{ppi}.

\item[7)] T-duality at $D = 9$ relates type IIA and type IIB
theories, as well as SO(32) and $E_8 \times E_8$ heterotic strings in
their broken phase SO(16)$\times$SO(16)~\cite{jji}.

\item[8)] M-theory and strings may undergo a further
unification in twelve dimensions ($F$-theory)~\cite{bbi}.

\item[9)] M-theory in the infinite momentum frame is equivalent
 to the large-$N$ limit of certain Yang--Mills theories \cite{bfss},
\cite{dvv},\cite{seib}.

\item[10)] Brane dynamics in M-theory reproduces non-perturbative aspects of
 rigid
super-Yang--Mills theories and in particular the Seiberg--Witten result 
\cite{wit}.
\end{itemize}

In the next sections we will describe some aspects of the above results,
namely some non-perturbative properties of superstring theories.
In particular in section 4 we will describe duality symmetries of M-theory in
various dimensions and their connections to S- and T-dualities of string
 theories,
in section 5 some particular aspects of the physics of extremal BPS 
black-holes,
in section 6 the Matrix-Model formulation of M-theory and finally in section 7
 an application
of these non-perturbative results to coupling-constant unification.


\section{Dualitites in Lower Dimensions}
A major role in exploring non-perturbative properties of
supersymmetric theories when compactified to dimensions $D<10$ is
played by the so-called U-duality group \cite{ht}.
This group combines the S-type duality \cite{esse}, {\it i.e.} the
 coupling-constant inversion,
 with the so-called T-duality, {\it i.e.} the
geometrical symmetries of the manifold of compactification, augmented
with some discrete translational symmetries, which appear in a
natural way in the perturbative string formulation \cite{hhia}.

For toroidally compactified strings, where the number of unbroken
supersymmetries does not decrease in the process of compactification,
these discrete groups are always subgroups of the non compact continuous
 duality
groups of the effective maximally-extended supergravity field theory \cite{cj}.

These groups play an important role in the non-perturbative study of
these theories.

Two important applications, discussed in the subsequent sections, are
the black-hole entropy of BPS states preserving one supersymmetry and
the Matrix-model formulation of M-theory.

For Type II string theory compactified on a $d$-dimensional torus $T^d$, the
U-duality group is $E_{d+1}(\ZZ)$, which is a discrete subgroup of the
maximal non compact form of the $E_{d+1}$ group.
For $d=1,2,3,4$, these groups are $GL(2,\IR)$, $SL(3,\IR) \times
SL(2,\IR)$, $SL(5,\IR)$, $O(5,5)$ respectively.
For $d=5,6$ we encounter the exceptional groups $E_{6(6)}$ and  $E_{7(7)}$.

Note that these groups have rank $d+1$ and, being maximally non
compact, their Cartan subalgebra can be taken to be $O(1,1)^{d+1}$.
These generators correspond to the string coupling and to the $d$
circles in $T^d\sim (S_1)^d$.

Under S--T-duality, the group $E_{d+1} \supset O(1,1)\times O(d,d)$ at $d<6$
( $E_{7} \supset SU(1,1)\times O(6,6)$ at $d=6$) \cite{noi}, \cite{lupost}.

The corresponding discrete subgroup is $ Z_2 \times O(d,d;\ZZ)$ at $d<6$ 
($ SL(2,\ZZ)\times O(6,6)$ at $d=6$), {\it
i.e.} the coupling-constant inversion and the T-duality of
ten-dimensional strings compactified on $T^d$.

This subgroup does not mix R-R and N-S states.
The remaining transformations in $E_{d+1}$ mix  R-R and N-S states
and generate U-dualities.
These transformations mix D-branes \cite{pol} with N-S branes and
correspondingly R-R charges with N-S charges.
These symmetries are non-perturbative in nature.

To understand the charge content of a generic multicharge
configuration, it suffices to decompose the charge vector of the
U-duality group with respect to S--T-dualities.

Let us consider, as an example, the $D=5$ and $D=4$ black-holes,
corresponding to 0-branes at $D=5$ and $D=4$.

The generic charge vector is in the ${\bf 27}$ of $E_{6(6)}$ at $D=5$
and  in the ${\bf 56}$ of $E_{7(7)}$ at $D=4$.

Its decomposition under S-T duality is:
\begin{eqnarray}
27 & \rightarrow & 16_{1} + 10_{-2} + 1_4 \nonumber\\
56 & \rightarrow & (1,32) + (12,2).
\end{eqnarray}
The R-R charges are $O(d,d)$ spinors, therefore there are  16
charges at $D=5$ and 32 (dyonic) charges at $D=4$.
These charges correspond to D-branes configurations wrapped and
reduced in different ways.


\section{Black-Hole Entropy and Bekenstein-Hawking  Formula}
The U-duality classification of black-holes, as well as of other
extended objects, plays a central role in the determination of the
entropy of a multicharge configuration.

The entropy can be computed either with a microscopic calculation,
using D-brane techniques, by counting microstates \cite{stva}, \cite{cama},
 or using the
Bekenstein--Hawking area formula \cite{bh} from the low-energy supergravity
effective field theory \cite{black}, 
\cite{cve}, \cite{fk}, \cite{45d}, \cite{malda}.

Since the entropy is topological in nature~\cite{lars} and does not depend
 on the
moduli fields, {\it i.e.} the  v.e.v. of scalar fields at
infinity, it can only depend on quantized charges.
Because of U-duality it must be a U-invariant combination of the
charges.

Indeed the black-hole entropy, at $D=5$ and $D=4$, is proportional to
the square root of the unique cubic and quartic invariants of the
fundamental representations of $E_{6(6)}$ and $E_{7(7)}$
respectively \cite{fk}, \cite{kako}.

This result can be obtained, in the supergravity framework, by computing
the horizon area of an extremal BPS black hole preserving $1/8$ of
supersymmetry.
The actual value is proportional to $M_{extr}^{3/2}$ at $D=5$ and to
$M_{extr}^{2}$ at $D=4$, where $M_{extr}$ is the value of the BPS
mass extremized in the moduli space \cite{fk}.
The extrema occur for moduli at rational values in terms of
 the quantized electric and magnetic charges \cite{fks}.
The latter is the BPS mass of the so-called double extreme
black holes \cite{ksw}.

To make contact with the D-brane microscopic calculation, it is useful
to compute the cubic and quartic invariants in the so-called
``normal frame'' \cite{fesazu}, {\it i.e.} in the frame where the charge 
vector is
represented by a skew diagonal matrix.

Here we will briefly describe the 5-$D$ case \cite{fema}.

At $D=5$, in terms of the three eigenvalues $e_i$ of a traceless $8\times
8$ matrix, the $27^3$ invariant reduces to:
\begin{equation}
I(27^3) = s_1s_2s_3
\end{equation}
where
\begin{equation}
e_1=s_1+s_2-s_3 \, , \quad  e_2=s_1-s_2+s_3 \, , \quad  e_3=-s_1+s_2+s_3
\end{equation}

Taking the case of Type II on $T^5$ we can choose $s_1$ to correspond
to a solitonic five-brane charge, $s_2$ to a fundamental string
winding charge along some direction and $s_3$ to Kaluza-Klein
momentum along the same direction.

We can see that in this specific example  one breaks $1/8$
supersymmetry if $s_1s_2s_3 \neq 0$,  $1/4$
 if $s_1s_2 \neq 0$, $s_3 =0$, $1/2$
 if $s_1 \neq 0$, $s_2=s_3=0$.

The case of enhanced supersymmetry corresponds to the invariant
combinations $I_3=0$ and $ \frac{\partial I_3}{\partial e_i}=0$ and
selects specific orbits of the ${\bf 27}$, {\it i.e.} the generic
orbit ($I_3 \neq 0$) preserving $1/8$ supersymmetry is
$E_{6(6)}/F_{4(4)}$, while the critical orbit, preserving $1/2$
supersymmetry, is  $E_{6(6)}/O(5,5) \otimes T_6$ \cite{fegu}, \cite{lps}.

The basis chosen in the  above example is S-dual to the D-brane basis
usually chosen for describing black holes in Type IIB on $T_5$.
All other bases are related by U-duality to this particular choice.
We also observe that the above analysis relates the cubic invariant to the
 picture  of 
intersecting branes since a three-charge $1/8$ BPS configuration with non 
vanishing entropy
  can be thought as obtained by intersecting three single charge $1/2$ BPS
 configurations
\cite{bdl}, \cite{bll}, \cite{lptx}

By using the S--T-duality decomposition we see that the cubic
invariant reduces to $I_3(27)=10_{-2} 10_{-2} 1_4 + 16_1 16_1
10_{-2}$.
The 16 correspond to D-brane charges, the 10 correspond to the 5 KK
directions and winding of wrapped fundamental strings, the 1
correspond to the N-S five-brane charge.

We see that to have a non-vanishing area we need a configuration with
three non-vanishing N-S charges or two D-brane charges and one N-S charge.

Unlike the 4-$D$ case, it is impossible to have a non-vanishing entropy
for a configuration only carrying D-brane charges.


\section{Matrix Model Formulation of M-Theory}
In the last year a proposal for a non-perturbative formulation of
M-theory  has been made  in which M-theory compactified on $T^d$ is
related to a $d+1$ supersymmetric Yang--Mills theory compactified on a
dual torus  $ \tilde T^d$ \cite{bfss}.

Since the Yang--Mills coupling $g^2$ in $d+1$ dimensions has
dimensions $m^{3-d}$, we may introduce a dimensionless running
coupling constant as $ \hat g^2(\mu) = g^2 \mu^{d-3}$.
In particular the coupling strength  at scales of the order of the torus size
$ \tilde L$  is  $ \hat g^2(\mu) = g^2  \tilde L^{3-d}$.

Then we can study the ultraviolet (infra-red) properties of the theory
by decreasing (increasing) $ \tilde L$.
For $d<3$, going to strong dimensionless coupling probes the infra-red
behaviour of super-Yang--Mills theory.
For $d>3$ it probes the ultraviolet.

Matrix theory is defined as the large $N$ limit of a $U(N)$
 super-Yang--Mills theory with the infinite momentum limit in which the
$11$-th direction is chosen to be the longitudinal direction.
The theory is assumed to be compactified in this direction on a circle
of size $R$.
When $d$ transverse directions are compactified on a $d$-torus then they will
 be
related to M-theory compactified on a dual torus.
In the original  discussion the starting point was the theory of D0-branes
in IIA theory when the compact direction is identified with the
longitudinal direction \cite{bfss}, \cite{dvv}.

The starting point is that M-theory on $S_1$ of size $L$ is
equivalent, through the Matrix-Model construction, to uncompactified
perturbative Type IIA string theory in the limit of shrinking torus size.

The limit $L \rightarrow 0$ corresponds to strongly coupled
Yang--Mills theory with $\hat g^2 = (2\pi)^4(\ell _{11}/  L)^3$  and
string coupling constant $g_s^2 = 2 \pi /\hat g^2$ ($\ell _{11}$ is
the 11-$D$ Planck length).

The relation between the dual radius $\tilde L$ on which Yang--Mills
theory is compactified and the M-theory radius $L$ is: $\frac{\tilde
L}{4\pi^2}= \frac{\ell _{11}^3}{RL}$ ($R$ is the radius of the
compactified longitudinal direction).

In particular the limit of strongly coupled Yang--Mills theory becomes
the weakly coupled IIA string theory in $D=10$.

In the case of the two torus, the U-duality group is $SL(2,\ZZ)\times
Z_2$, and $SL(2,\ZZ)$ corresponds to the geometrical symmetry of the
two-torus.

In the case of the three-torus we have 4-$D$ Yang--Mills theory.
The U-duality group is $SL(3,\ZZ)\times SL(2,\ZZ)$,  where the first
factor can be interpreted as the geometrical symmetry  and the second
factor to the electric-magnetic duality of four dimensional
 super-Yang--Mills theory.

The analysis becomes more subtle when $d>3$.
First, the Yang--Mills theory becomes non renormalizable.
Moreover the electric and
magnetic fluxes of super-Yang--Mills theory do not complete the
U-duality multiplets; however proposals have recently been made for the cases
$d=4$ and $d=5$ in terms of fixed-point six-dimensional $(2,0)$
superconformal field theories  and of N-S five-branes \cite{seib}.


\section{Gauge Couplings Unification}

New predictions of  non-perturbative string theories can be derived from
these non perturbative relations between the five  seemingly  different
superstring theories.  As a circumstantial example \cite{qqi},
strongly coupled heterotic string meets the agreement of $\alpha_{GUT}$ as
measured (at LEP) from low-energy data.

In weakly coupled heterotic string, compactified on a Calabi--Yau threefold
 of size
$V \approx M_{GUT}^{-6}$ with $G_N = \frac{e^{2\phi}(\alpha')}{64\pi V}$~,
\[
~\alpha_{GUT} = e^{2\phi}(\alpha')^3/16\pi V \rightarrow G_N =
\alpha_{GUT}\alpha'/4~.
\]
If $e^{2\phi} \leq 1, G_N \geq \alpha_{GUT}^{4/3}/M^2_{GUT}$, which is too 
large
with respect to experiment.

In type I string (weak coupling)~,
\begin{eqnarray}
\alpha_{GUT} &= &\frac{e^{\phi_I}(\alpha')^3}{16\pi V},~G_N =
\frac{e^{2\phi_I}(\alpha')}{64 \pi V}\nonumber \\
 &\rightarrow& G_N = e^{\phi_I}\alpha_{GUT}
\alpha'/4~.\nonumber
\end{eqnarray}
Here, $G_N$ can be small.

  In the M-theory set-up ($\kappa$ is the 11-$D$ gravitational coupling and
 $\rho$
the compactification radius):
\[
G_N = \frac{k^2}{16\pi^2V\rho}~,~~\alpha_{GUT} =
 \frac{(4\pi k^2)^{2/3}}{2V}\ll
1~.
\]
So no disagreement with the experimental input exists in principle.

Finally, supersymmetry breaking can be described in a natural way both through
 a
strongly coupled hidden gauge sector leading to gaugino
 condensation~\cite{rri}, \cite{xxi} and
through the no-scale structure~\cite{ssi} of M-theory.  The decompactification
problem may be avoided~\cite{tti}--\cite{vvi}.


\section*{Conclusions}
In this status report we have summarized the tremendous advances that have
 been made 
over the last three years in our understanding of non-perturbative
phenomena of quantum theories encompassing the fundamental interactions.
These results heavily used mathematical concepts such as algebraic geometry
 \cite{greene},
space-time symmetries such as supersymmetry, and topological configurations
such as BPS solitonic states of different nature \cite{duff}. 
The old idea of electric--magnetic duality has revealed new insights in the
 dynamics of these theories
with major advances achieved between the Paris and Brisbane ICMP conferences.
\par
We hope that further progress and new directions, such as the understandig of
 the origin
of symmetry breaking and the real spectrum of elementary particles, can be
 conceived 
in the years to come.

It is plausible that even more stunning developments will be reported at the
 Fourteenth
ICMP Conference, to be held in Postdam at the end of this millenium.


\end{document}